\documentclass[twoside,showpacs,superscriptaddress,twocolumn,floatfix,a4paper,pra]{revtex4-1}

\usepackage{hyperref}
\usepackage{graphicx}
\usepackage[utf8x]{inputenc}
\usepackage[T1]{fontenc}
\usepackage{siunitx,  multirow} %, amsmath,}
\usepackage{marvosym}

\usepackage[rgb]{xcolor}
\begin{document}

\title{Experimental tests of coherence and entanglement conservation}

\author{Antonín Černoch} %\email{antonin.cernoch@upol.cz}
\affiliation{RCPTM, Joint Laboratory of Optics of Palacký University and Institute of Physics of Czech Academy of Sciences, 17. listopadu 12, 771 46 Olomouc, Czech Republic}

\author{Karol Bartkiewicz} %\email{bark@amu.edu.pl}
\affiliation{Faculty of Physics, Adam Mickiewicz University, PL-61-614 Pozna\'n, Poland}
\affiliation{RCPTM, Joint Laboratory of Optics of Palacký University and Institute of Physics of Czech Academy of Sciences, 17. listopadu 12, 771 46 Olomouc, Czech Republic}

\author{Karel Lemr} %\email{k.lemr@upol.cz}
\affiliation{RCPTM, Joint Laboratory of Optics of Palacký University and Institute of Physics of Czech Academy of Sciences, 17. listopadu 12, 771 46 Olomouc, Czech Republic}

\author{Jan Soubusta}
\affiliation{Institute of Physics of Czech Academy of Sciences, Joint Laboratory of Optics of PU and IP AS CR, 17. listopadu 50A, 772 07 Olomouc, Czech Republic}

\date{\today}

\begin{abstract}

We experimentally demonstrate the migration of coherence between composite quantum systems
and their subsystems. The quantum systems are implemented using polarization states of photons
in two experimental setups. The first setup is based on linear optical controlled-phase quantum 
gate and the second scheme is utilizing effects of nonlinear optics. 
Our experiment allows to verify the relation between correlations of the subsystems
and the coherence of the composite system, which was given in terms
of a conservation law for maximal accessible coherence by Svozilík \emph{et al.}  
[Phys. Rev. Lett. {\bf 115}, 220501 (2015)].
We observe that the maximal accessible coherence is conserved for the implemented class of global evolutions of the composite system.
\end{abstract}


\maketitle

% -------------------------------------------------------------------------------------------------
\section{Introduction}

Fundamental laws in physics can often be formulated in terms of the conservation principles.
These laws not only represent our understanding of the underlying physical phenomena but can also be used to predict time evolution of quantum correlations in the investigated systems \cite{Lostaglio_PRX5, Cwiklinski_PRL115}.

Traditionally, the quantities conserved in the closed systems include overall energy or momentum.
These conservation laws can be described even within the confines of classical mechanics. 
Quantum theory provides conservation laws for some additional, more abstract quantities. 
For example, it has been established by Englert, Greenberger and Yasin that the sum of interference pattern visibility and the coefficient of determination of the trajectory of a particle is also a conserved quantity \cite{Englert_PRL77,Greenberger_PLA128}. 
This conservation law describes quantitatively our fundamental insight into wave-particle duality, i.e., the more we know about particle’s trajectory, the less it would manifest its wave properties -- namely the interference 
\cite{Englert_PRL77,Greenberger_PLA128,Bjork_PRA58, Bjork_PRA60, Jakob_OC283, Streltsov_RevModPhys89, Davidovich_PRL98,Jaeger_PRA51}

Coherence is a consequence of the principle of superposition which is a key property of quantum states.
It typically manifests as interference patterns which have been observed in various physical systems including both strong and ultraweak optical fields
\cite{kocsis2011observing}, electrons \cite{davisson1928reflection,frabboni2007young}, 
atoms or even fullerene molecules \cite{arndt1999wave}. 

While coherence is also well known in classical physics, in quantum physics, 
this concept is further broaden to the coherence between two or more distinct parties also known as the entanglement \cite{HorodeckiRMP}. 
The coherence of nonclassical light (in terms of photon statistics) 
can be expressed as its ability to create entanglement~\cite{Miranowicz_PRA92}.
Recently, theoretical research let to formulation of other measures of coherence that are invariant under coherence preserving evolutions \cite{Levi_NJP16,Baumgratz_PRL113,Streltsov_PRL115,Chan_PRA75}. 
In 2015, Svozilík {\it et al.} 
put forward the conservation law for the maximum accessible coherence under global unitary evolutions. 
They have shown how coherence migrates in multipartite quantum systems from the classical coherence of a given subsystem to the quantum correlations between subsystems \cite{Svozilik_PRL115}. 
In about the same time, other research groups investigated the coherence migration 
for Gaussian states \cite{Ge_PRA92}, single-photon states \cite{Miranowicz_PRA92} 
and subsequently in general \cite{Killoran_PRL116}.

Previously-mentioned theoretical results on coherence migration provide valuable insight into the relation between classical coherence and quantum correlations. So far, however, these results were not subjected to experimental verification. In this paper we report on such a test. 
We verify the conservation of the maximally accessible coherence while it migrates
between classical coherence and quantum correlations as described in Ref \cite{Svozilik_PRL115}. 
For this purpose, we have selected two optical processes described by global unitary evolution. 
The first process, presented in Sec.~\ref{SEC:setup1}, involves a linear-optical controlled-phase (c-phase) gate. 
In the second experiment, we observe entanglement generation in the process of spontaneous parametric down-conversion from a partially coherent pump beam (see Sec.~\ref{SEC:setup2}). 
Finally in Sec.~\ref{SEC:discussion} we discuss the results and conclude.

% -------------------------------------------------------------------------------------------------
\section{Coherence migration under c-phase operation}\label{SEC:setup1}

\begin{figure}
  \includegraphics[width=.8\columnwidth]{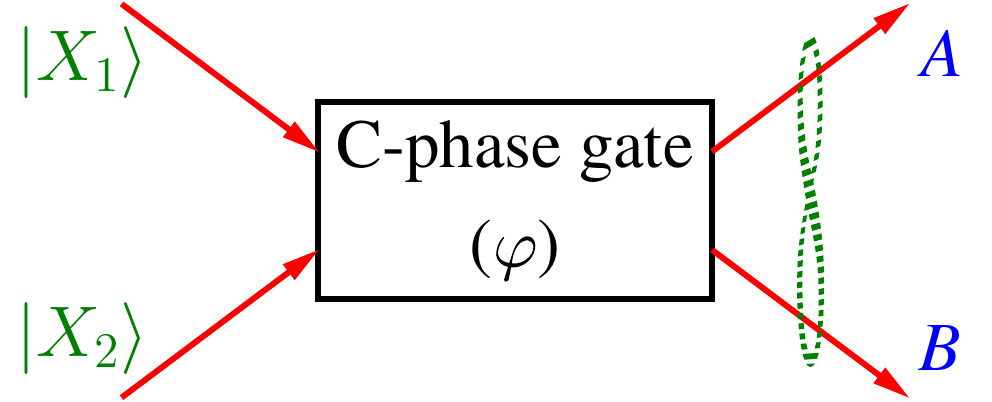} %{setup_c-phase.eps}
  \caption{(color online) Conceptual scheme of the experimental setup with c-phase gate adjusted at phase $\varphi$.
           \label{FIG:setup1}}
\end{figure}

First, we analyze the migration of accessible coherence between classical coherence and quantum correlations in the experimental setup based on a tunable linear optical c-phase gate \cite{Lemr_PRL106, Lemr_PRLA86}. 
This setup is schematically depicted in Fig.~\ref{FIG:setup1} and described in detail in Ref.~\cite{Lemr_PRL106}. 
The c-phase gate performs the unitary input-output transformation:
\begin{eqnarray}
    |H H \rangle_{1,2} \to |H H \rangle_{A,B},\qquad & |H V \rangle_{1,2} \to |H V \rangle_{A,B} , \nonumber \\
    |V H \rangle_{1,2} \to |V H \rangle_{A,B},\qquad & |V V \rangle_{1,2} \to e^{i \varphi}|V V \rangle_{A,B},
  \label{EQ:U_C-phase}
\end{eqnarray}
where the single-photon basis polarization states are horizontal ($|H \rangle$)  and 
vertical ($|V\rangle$) linear polarizations.
The phase $\varphi$ is a tunable parameter of this quantum gate. 
Qubits are encoded as polarization states of single photons. Thus, 
the classical coherence corresponds to the degree of polarization.

A separable two-photon state $|++ \rangle$, where $|+ \rangle = (|H\rangle+|V\rangle)/\sqrt{2}$
denotes diagonal linear polarization, is inserted at the input of the gate. 
Coherence of each photonic qubit can be calculated from its reduced density matrix 
$\hat{\rho}_{i}$ for $i=A,B$. Knowing the density matrix of the entire system $\hat{\rho}$, 
we can calculate the reduced matrices of individual photons as a partial trace,
$\hat{\rho}_A = \mathrm{Tr}_{B}[\hat{\rho}]$ for the first and
$\hat{\rho}_B = \mathrm{Tr}_{A}[\hat{\rho}]$ for the second photon. 

Coherence proportional to the radius of a Poincaré sphere of each subsystem is calculated as
\begin{equation}
  D_i^2 =  {\rm Tr}[\hat{\rho}_i^2] - \frac{1}{2}, \qquad i = {A,B}, \label{eq:D}
\end{equation}
where we use the notation identical to Ref. \cite{Svozilik_PRL115} up to a scaling factor of $\sqrt{2}$.
We introduce a mean coherence of the two photons, $D^2 = (D_A^2+D_B^2)/2$. 
For the considered input state $|++\rangle$, we have $D_A^2 = D_B^2 = D^2 = \frac{1}{2}$.

Mutual correlations between the two photons can be quantified by $T^2$~\cite{Svozilik_PRL115} given as
\begin{equation}
  T^2 = {1 \over 4}
      \left( 
       1 + \sum_{i,j = 1}^3{t_{ij}^2} 
      \right),
  \quad 
  t_{ij} = {\rm Tr}[\hat{\rho} \hat{\sigma}_i \otimes \hat{\sigma}_j],\label{eq:T}
\end{equation}
where $\hat{\sigma}_i (i = 1,2,3)$ are Pauli matrices. 

This quantity relates to various two-qubit entanglement witnesses 
based on eigenvalues of the Horodecki matrix $R_{ij}=t^2_{ij}$ (see Refs.~\cite{HorodeckiBIV,BellFEF17}
and references therein), as $\mathrm{Tr}R=4T^2-1$. 
In particular, the maximal Bell-Clauser-Horne-Shimony-Holt quantity \cite{CHSH,HorodeckiBIV} 
can be expressed as $B=2\sqrt{\mathrm{Tr}R-\min{}{}[\mathrm{eig}(R)]}$,
$B\le 2$ for spatially separated classical two-level systems and $B\le2\sqrt{2}$
for quantum systems. This quantity can be used for measuring 
the degree of nonlocality and to estimate various measures of 
quantum entanglement~\cite{HorstBIV,BellFEF17}. 
It follows from $\min{}{}[\mathrm{eig}(R)]\le \frac{1}{3}\mathrm{Tr}R$ that $B\ge 2\sqrt{2\mathrm{Tr}R/3}=2\sqrt{(8T^2-2)/3}$. Hence, we witness nonlocality and entanglement ($B>2$) if $T^2>5/8=0.625$.

\begin{table}
  \caption{Parameters of the two-photon state at the output of the c-phase gate for seven 
           values of the phase $\varphi$. $D_A, D_B$ -- local coherence (degree of 
           polarization) of individual photons, $T^2$ -- degree of correlation, 
           $S^2$ -- maximal accessible coherence.
           \label{TAB:result1}}
 \begin{tabular}{l cccccccc} 
  \hline \hline
   $\varphi/\pi$ & $D_A$ & $\delta D_A$ & $D_B$ & $\delta D_B$ & $T^2$ & $\delta T^2$ & $S^2$ & $\delta S^2$ \\ 
  \hline 
   0     & 0.707 & 0.000 & 0.707 & 0.000 & 0.500 & 0.000 & 1.000 & 0.000 \\
   0.05  & 0.695 & 0.003 & 0.701 & 0.001 & 0.502 & 0.002 & 0.989 & 0.004 \\
   0.125 & 0.682 & 0.002 & 0.681 & 0.002 & 0.532 & 0.002 & 0.996 & 0.002 \\
   0.25  & 0.648 & 0.004 & 0.652 & 0.003 & 0.554 & 0.003 & 0.976 & 0.004 \\
   0.5   & 0.619 & 0.003 & 0.607 & 0.004 & 0.599 & 0.005 & 0.974 & 0.005 \\
   0.75  & 0.521 & 0.004 & 0.518 & 0.005 & 0.701 & 0.006 & 0.971 & 0.005 \\
   1     & 0.359 & 0.006 & 0.330 & 0.007 & 0.825 & 0.007 & 0.944 & 0.007 \\
  \hline \hline
 \end{tabular}
\end{table}

\begin{figure}
  \includegraphics[width=\columnwidth]{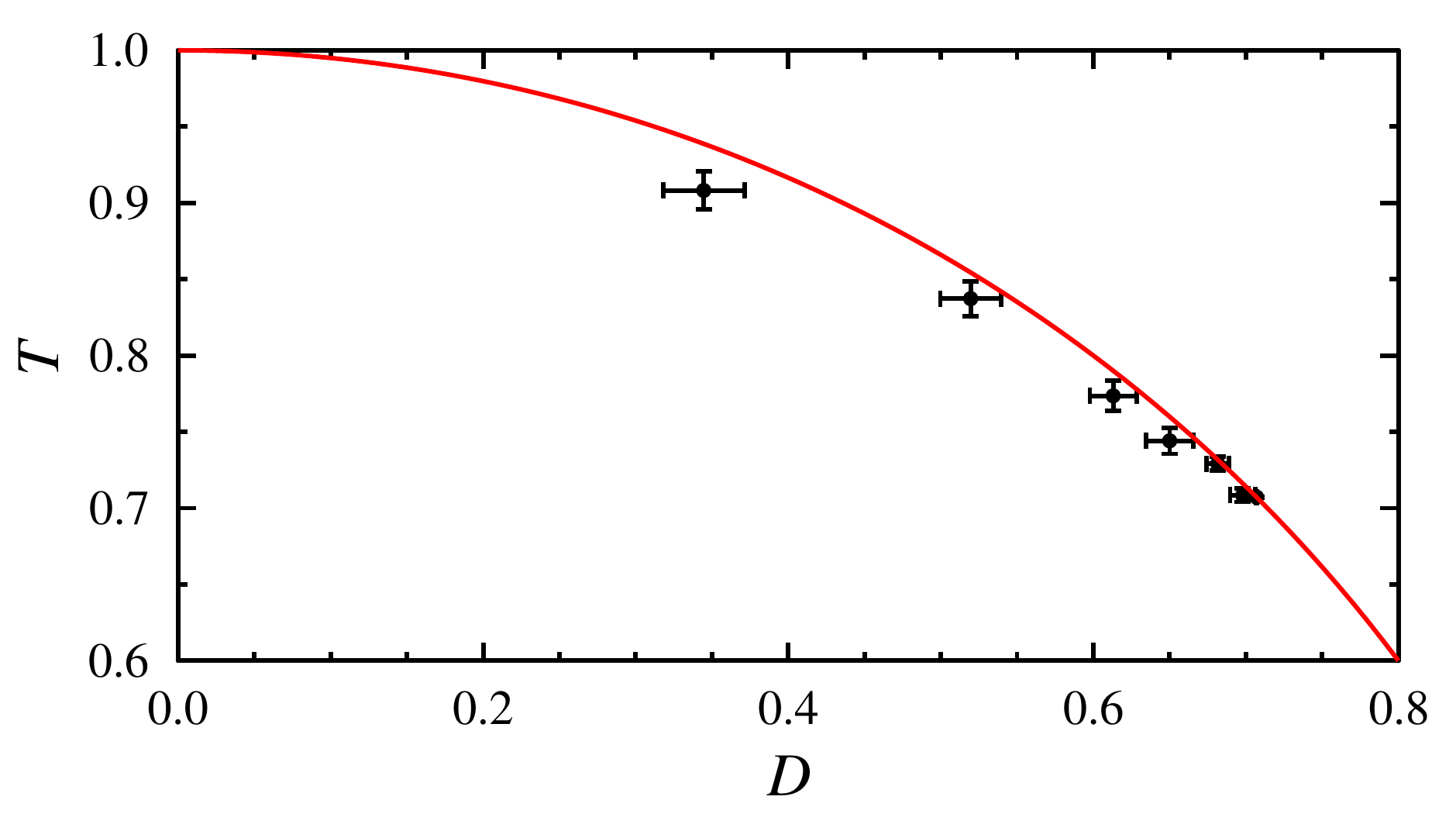}
  \caption{(color online) Parameters of the two-photon state at the output of the c-phase gate as a function of the phase $\varphi$. 
            See text for definitions of quantities $T$ and $D$.
            Red curve represents theoretical dependance according to Eq.~(\ref{eq:TD}). 
           \label{FIG:result1}}
\end{figure}

The degree of correlation $T^2$ of the separable input state reaches
its minimum value of $1/2$. The c-phase gate is however capable 
to continuously change (in this case increase) the 
entanglement of the two-photon state from the minimum value of $T^2 = 1/2$ for a separable state to 
$T^2=1$ corresponding to a maximally entangled Bell state. This is achieved by tuning the phase $\varphi$ in the range from 0 to $\pi$.

In this experiment the classical coherence of the two-photon input state partially migrates to the degree of 
correlation $T^2$ at the output. In this process the conserved quantity is called the maximal accessible coherence $S^2$ of the two-photon system,
\begin{equation}
  S^2 = D^2 + T^2. \label{eq:TD}
\end{equation}
The value of the conserved parameter $S^2$  equals to
the purity  $P = \mathrm{Tr}[\hat{\rho}^2]$ of the composite system~\cite{Svozilik_PRL115}. 
Ideally, purity of the two-photon state shall be equal to 1. 
Due to experimental imperfections, the observed output state purity typically fluctuates between 0.9 and 1.

The output state density matrix was reconstructed with optimal  quantum state tomography and maximum likelihood estimation \cite{srep19610,PhysRevA.90.062123}. This has been performed for seven phase shifts $\varphi$. Each time, we have calculated the values of $T^2$ and $D^2$ using the above mentioned formulas. We have also developed a theoretical model assuming perfect c-phase operation. This way we can predict the ratio of coherence migrated from the input state into the output state entanglement as the phase shift $\varphi$ is scanned from 0 to $\pi$. To compare our data to the ideal theoretical model, the values of $D^2$ and $T^2$ need to be normalized to the overall output state purity. The experimentally obtained values of $T^2$, $D^2$ and $S^2$ are summarized in Tab.~\ref{TAB:result1}. In Fig.~\ref{FIG:result1} we visualize both theoretical and experimental ratio of coherence migrated into entanglement. Up to experimental uncertainty, our data are in accordance with the theoretical prediction.

%-------------------------------------------------------------------------------------------------
\section{Coherence migration in a nonlinear optical proccess}\label{SEC:setup2}

\begin{figure}
  \includegraphics[width=\columnwidth]{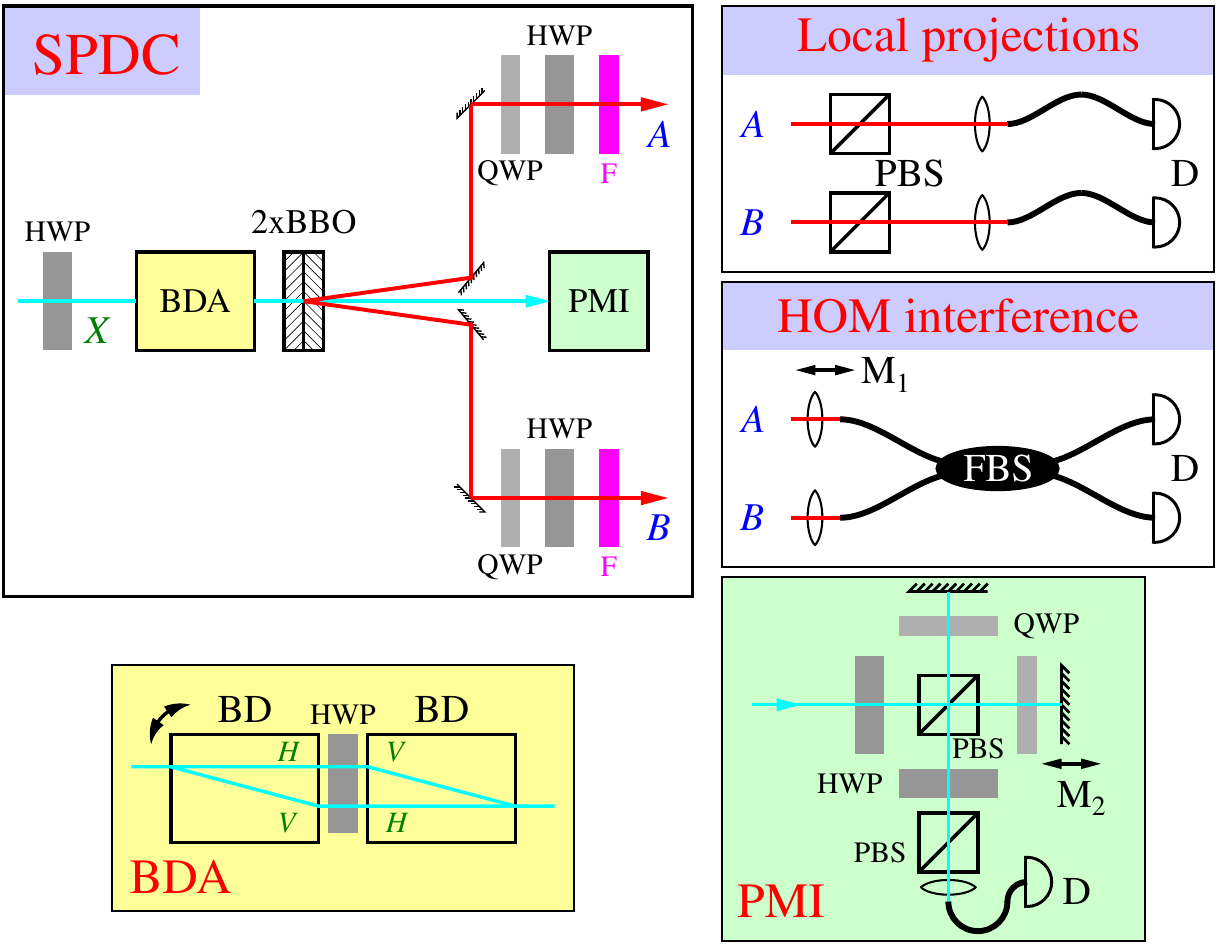}
  \caption{(color online) Scheme of the second experimental setup with SPDC
           generation of two-photon states. HWP -- half-wave plate, QWP -- quarter-wave plate,
           F -- spectral filter, BDA -- beam displacer assembly, BD -- beam displacer,
           PMI -- polarizing Michelson interferometer, 
           PBS -- polarizing beam-splitter, FBS -- fiber beam-splitter, D -- detector, 
           M -- motor translation.
           \label{FIG:setup2}}
\end{figure}

The second setup is a typical scheme for generation of two-photon states in the process of spontaneous parametric down-conversion (SPDC) in a nonlinear material. We use cascade of two BBO crystals also called the Kwiat configuration \cite{Kwiat_PRA60, White_PRL83, Halenkova_AO51}.
The optical axis of the first (second) crystal is in the vertical (horizontal) plane.
This perpendicular configuration of two Type I crystals allows direct generation of photon pairs 
correlated in polarization. Scheme of the second setup is drawn in Fig.~\ref{FIG:setup2}.
Here the input coherence is represented by coherent superposition of $H$ and $V$ polarization components of the pump beam. 
When these two polarization components are mutually delayed using a beam displacer assembly (BDA), the coherence of the pump beam is decreased. 
To inspect the value of the displacement we subject the pumping beam to a polarization Michelson interferometer (PMI) placed behind the crystal cascade (see Fig.~\ref{FIG:setup2}). 
The value of the displacement is inferred from the position of autocorrelation function maxima.

The horizontal component of the pump beam leads to spontaneous parametric generation of a vertically polarized pair of photons $|VV\rangle$. Similarly, the vertically polarized pump beam is downconverted into a pair of horizontally polarized photons $|HH\rangle$. When the horizontal and vertical pump beam components are temporarily overlapping, the overall two photon state generated by the crystals is a coherent superposition of the  $|HH\rangle$ and $|VV\rangle$ states, aka. the Bell state $|\Phi^+\rangle = (|HH \rangle + |VV \rangle)/\sqrt{2}$. It should be emphasized, that the case of highest down-conversion coherence corresponds to exact time overlap 
of the $H$ and $V$ components of the pump beam at the interface between the two BBO crystals.
In this case, the observed delay before or after the crystal cascade is 84 $\mu$m  due to the birefringence 
of the BBO crystals. The value of 84 $\mu$m represents a constant bias to be subtracted in subsequent analysis.

The broad spectrum of the SPCD photons is narrowed using band-pass filters F (FWHM of 3 nm). Fitting the experimental autocorrelation function of two-photon coincidences with a Gaussian curve yields
\begin{equation} 
  {\cal V} (d) = 0.029 + 0.945\, {\rm e}^{-(d/\sigma)^2},
\end{equation} 
where $d$ denotes the spatial displacement between the horizontal and vertical pump beam components. The corresponding FWHM is found to be FWHM = 2$\sqrt{\ln{2}} \sigma$ = \SI{142}{\micro\meter}. The coherence is proportional to value of this function in distance $d$ out of the maxima,
\begin{equation}
  S^2_{\rm in} = {\rm Tr} [\hat{\rho}_{\rm pump}^2] = (1- {\cal V}^2)/2. 
\end{equation}
In the experiment, we have selected ten specific values of polarization components displacement of the pump beam $d$. 

\begin{figure}
  \includegraphics[width=\columnwidth]{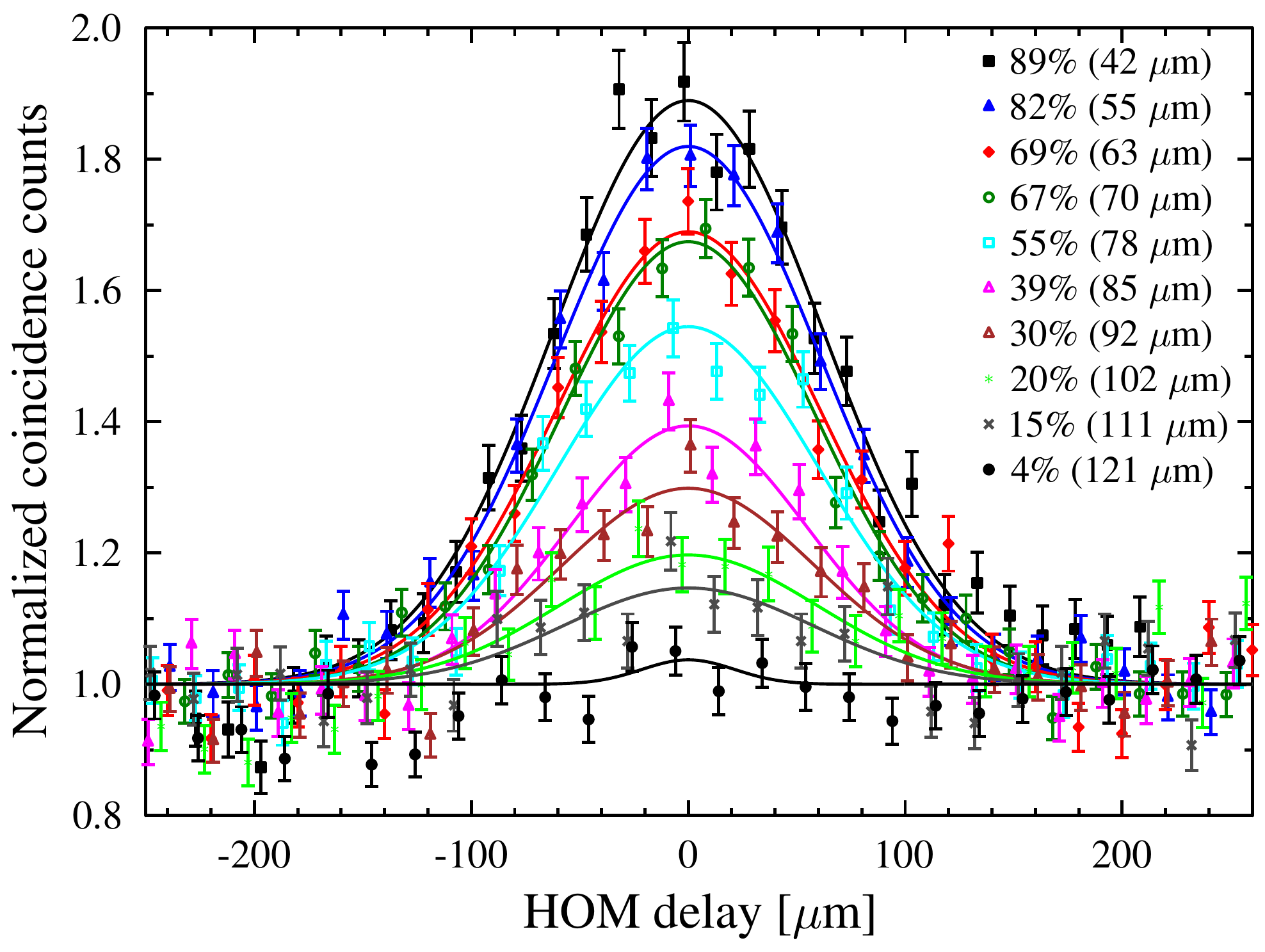}
  \caption{(color online) Antidips measured for the two-photon states starting from 
           $|\Psi^- \rangle$ for increasing delay of the $H$ and $V$ component of 
           the pump beam. The key shows the visibility for a given delay for all 		           measured curves.
           \label{FIG:antidip}}
\end{figure}

Simultaneously, the generated pairs of photons were collected into single-mode fibers leading to a balanced fiber coupler. Polarization state of the generated pair was transformed so that the two photons impinging on this fiber coupler are in a singlet-like state $|\Psi^{-}_d\rangle = \frac{1}{\sqrt{2}}\left(|HV\rangle - |VH\rangle_d\right)$, where the subscript $d$ denotes that the state is displaced. The displacement between the two-photon state components can not be resolved and the resulting two-photon state has to be described in the form of a mixed state $\hat\rho = p|\Psi^-\rangle\langle\Psi^-| + (1-p)|\Psi^+\rangle\langle\Psi^+|$, where $|\Psi^{\pm}\rangle = \frac{1}{\sqrt{2}}\left(|HV\rangle \pm |VH\rangle\right)$ are two of the Bell states. 

The value of $p$ can easily be measured as the probability of the two-photon state antibunching on the beam splitter (observing coincident detections behind the beam splitter, see Fig.~\ref{FIG:antidip}). Note that only photons in the singlet state $|\Psi^-\rangle$ deterministically antibunch while photons in any of the triplet Bell state bunch on the beam splitter. Measuring $p$ allows one to reconstruct the density matrix $\hat\rho$ and obtain the values of $T^2$ as defined in Sec.~\ref{SEC:setup1}. During the measurement we also tested the degree of polarization of individual photons 
from the SPDC pairs and found that $D^2$ is negligible ($\sim 4\times 10^{-3}$). To compensate for any experimental imperfections, we have measured the bunching effect on the beam splitter for a triplet state $|\Phi^+\rangle$. For this state, the visibility reaches 94\% which we use to correct the observed values of $p$. The output state coherence $S^2_\mathrm{out}$ was calculated from $D^2$ and $T^2$ using Eq.~(\ref{eq:TD}).

\begin{figure}
  \includegraphics[width=\columnwidth]{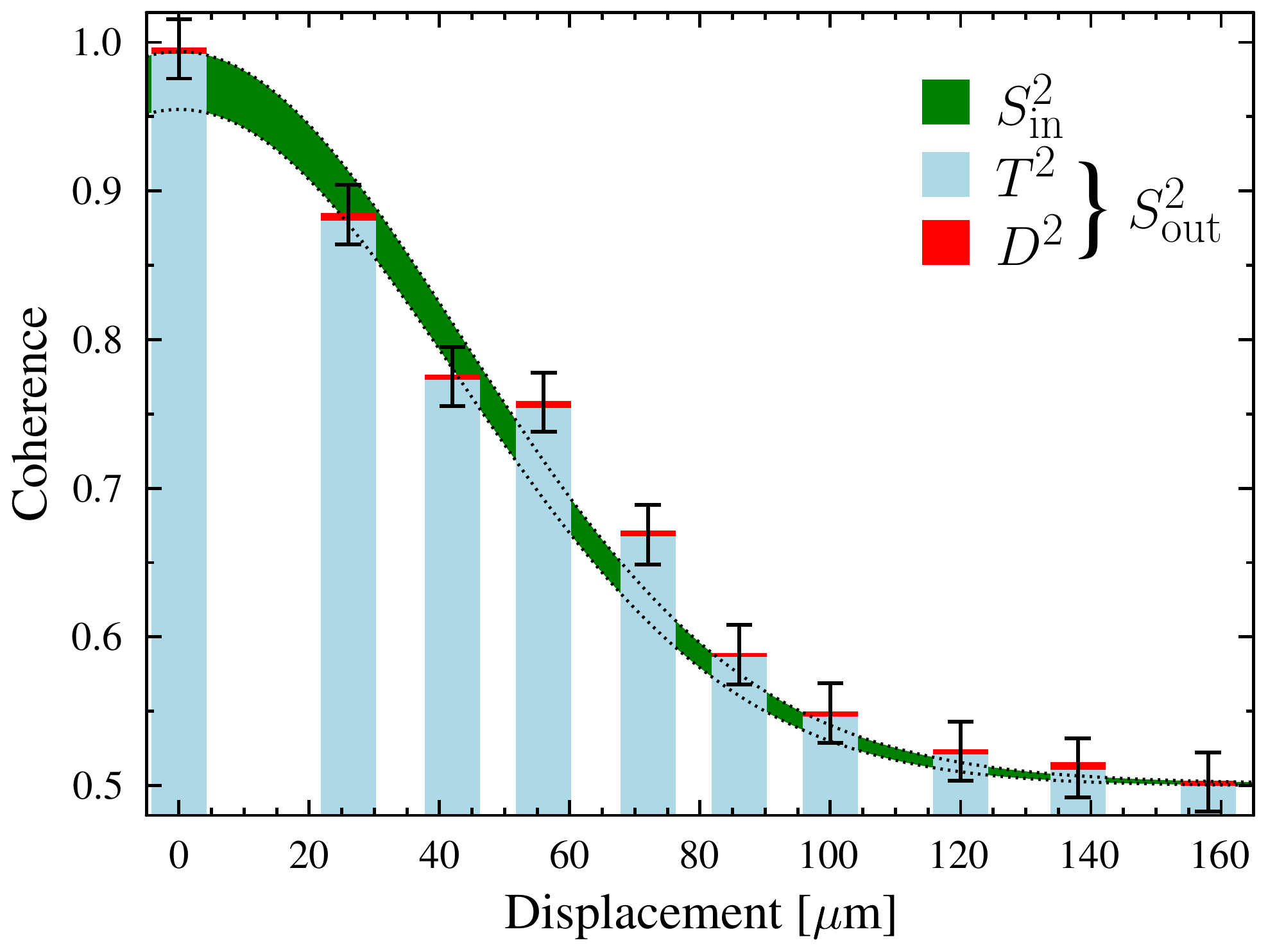}
  \caption{(color online) Coherence of the two-photon state generated by the Kwiat type crystal
           as a function of the displacement between the $H$ and $V$ component of the pump beam. Input coherence 
           $S^2_{\rm out} = D^2+T^2$ depicted using bars is compared with the input coherence 
           $S^2_{\rm in}$ depicted using green belt. Thickness of the belt represents measurement errors.
           \label{FIG:result2}}
\end{figure}

 We summarized the 
results in Tab.~\ref{TAB:result2} and in Fig.~\ref{FIG:result2}. It is clear that in this case 
of nonlinear interaction, the input maximal accessible coherence is carryed by the polarization 
of the pump beam. There is no entanglement at the input of the nonlinear crystals. At the output 
the generated photons seem unpolarized, corresponding to nearly zero degree of classical coherence. 
The whole input coherence of the pump is transferred to the degree of correlation between the SPDC-generated photons.

\begin{table}
  \caption{Parameters of the pump beam and SPDC-generated photons.
           $D_A, D_B$ -- local coherence (degree of polarization) of SPDC photons, 
           $T^2$ -- degree of correlation, $S^2_{\rm out}$ -- maximal accessible coherence of SPDC 
           photons, $S^2_{\rm in}$ -- maximal accessible coherence of the pump beam. The last
           row denoted Err shows typical error of the parameter in corresponding column.
            \label{TAB:result2}}
 \begin{tabular}{l cccccc} 
  \hline \hline
   d & $D_A$ & $D_B$ & $T^2$ & $S^2_{\rm out}$ & $S^2_{\rm in}$ & $|S^2_{\rm out}-S^2_{\rm in}|$ \\ 
  \hline
     0 & 0.064 & 0.023 & 0.993 & 0.996 & 0.974 & 0.021 \\
    26 & 0.069 & 0.033 & 0.881 & 0.884 & 0.896 & 0.012 \\
    42 & 0.030 & 0.041 & 0.774 & 0.775 & 0.796 & 0.021 \\
    56 & 0.074 & 0.043 & 0.755 & 0.758 & 0.706 & 0.052 \\
    72 & 0.044 & 0.040 & 0.669 & 0.671 & 0.620 & 0.050 \\
    86 & 0.015 & 0.032 & 0.588 & 0.588 & 0.568 & 0.020 \\
   100 & 0.037 & 0.034 & 0.548 & 0.549 & 0.535 & 0.014 \\
   120 & 0.010 & 0.049 & 0.522 & 0.523 & 0.512 & 0.011 \\
   138 & 0.041 & 0.061 & 0.512 & 0.515 & 0.505 & 0.010 \\
   158 & 0.048 & 0.030 & 0.501 & 0.502 & 0.502 & 0.001 \\
  \hline
   Err & 0.013 & 0.011 & 0.020 & 0.020 & 0.010 & 0.022 \\ 
  \hline \hline
 \end{tabular}
\end{table}

% --------------------------------------------------------------------------------------------------
\section{Conclusions}\label{SEC:discussion}

In this paper, we have demonstrated transformation between coherence and correlations in two optical experiments. 
In the first experiment a linear-optical c-phase gate is used to 
tune the transfer of individual coherence of the input photons into their correlations. 
We have verified that during this operation the maximal accessible coherence $S^2$ is conserved.

In the second experiment we tune the coherence of the pump beam which is generating 
two-photon states in the nonlinear SPDC process. The degree of correlation of the generated states 
is limited by the coherence of the original pump beam. The maximal accessible coherence 
$S^2$ is conserved also in this process. 

Our results experimentally prove that the conservation of accessible coherence over unitary operations belong to a broader set of conservations laws that are invaluable in predicting outcomes of complex physical phenomena.

\section*{Acknowledgements}
The authors acknowledge support by the Czech Science Foundation (Grant No. 
17-10003S) and the project 
No. CZ.02.1.01/0.0/0.0/16$\_$019/0000754 of the Ministry of Education, Youth and Sports of the Czech Republic.

% -------------------------------------------------------------------------------------- %

\end{document}